# Quantifying interictal intracranial EEG to predict focal epilepsy

Ryan S Gallagher[1,2*], Nishant Sinha[1,3,4*], Akash R. Pattnaik[1,3], William K.S. Ojemann[1,3], Alfredo Lucas[1,2,3], Joshua J. LaRocque[1,3,4], John M Bernabei[1,2,3], Adam S Greenblatt[4], Elizabeth M Sweeney[5], H Isaac Chen[6,7], Kathryn A Davis[1,4], Erin C Conrad[1,4**], Brian Litt[1,4**]

*, ** *denotes equal contribution as the first and senior authors.*

1) Center for Neuroengineering and Therapeutics, University of Pennsylvania
2) Perelman School of Medicine, University of Pennsylvania
3) Department of Bioengineering, University of Pennsylvania
4) Department of Neurology, University of Pennsylvania
5) Department of Biostatistics, Epidemiology and Informatics, University of Pennsylvania
6) Department of Neurosurgery, University of Pennsylvania
7) Corporal Michael J. Crescenz Veterans Affairs Medical Center

*Corresponding Authors:*
Ryan Gallagher: ryan.gallagher@pennmedicine.upenn.edu
Nishant Sinha: nishants@seas.upenn.edu

**Abstract**

**Introduction:**
Intracranial EEG (IEEG) is used for 2 main purposes, to determine: (1) if epileptic networks are amenable to focal treatment and (2) where to intervene. Currently these questions are answered qualitatively and sometimes differently across centers. There is a need for objective, standardized methods to guide surgical decision making and to enable large scale data analysis across centers and prospective clinical trials.

**Methods:** We analyzed interictal data from 101 patients with drug resistant epilepsy who underwent presurgical evaluation with IEEG. We chose interictal data because of its potential to reduce the morbidity and cost associated with ictal recording. 65 patients had unifocal seizure onset on IEEG, and 36 were non-focal or multi-focal. We quantified the spatial dispersion of implanted electrodes and interictal IEEG abnormalities for each patient. We compared these measures against the "5 Sense Score (5SS)," a pre-implant estimate of the likelihood of focal seizure onset, and assessed their ability to predict the clinicians' choice of therapeutic intervention and the patient outcome.

**Results:** The spatial dispersion of IEEG electrodes predicted network focality with precision similar to the 5SS (AUC = 0.67), indicating that electrode placement accurately reflected pre-implant information. A cross-validated model combining the 5SS and the spatial dispersion of interictal IEEG abnormalities significantly improved this prediction (AUC = 0.79; $p<0.05$). The combined model predicted ultimate treatment strategy (surgery vs. device) with an AUC of 0.81 and post-surgical outcome at 2 years with an AUC of 0.70. The 5SS, interictal IEEG, and electrode placement were not correlated and provided complementary information.

**Conclusions:** Quantitative, interictal IEEG significantly improved upon pre-implant estimates of network focality and predicted treatment with precision approaching that of clinical experts. We present this study as an important step in building standardized, quantitative tools to guide epilepsy surgery.

## Introduction

Over 20 million people worldwide have drug-resistant epilepsy, in which seizures are resistant to medical therapy.[1,2] Surgery can eliminate or markedly reduce seizures in these patients, but unfortunately 40-60% of patients who undergo surgery do not become seizure-free.[3–6] Modest outcomes from epilepsy surgery and variability in treatment across centers are driving researchers to develop rigorous quantitative methods and biomarkers to guide invasive treatment.[7–9] Prior work has focused on several major questions: *Is there a single focal seizure generating region, or is one present even if we haven't located it? What is the location of the generator(s)? What are the dynamics, rate, and pattern of seizure spread? and Where and how should we intervene?*[10–17] In this study we concentrate on the basic question of *if* there is a single focal seizure generator and whether therapy should be focal, such as surgical resection or ablation, or less targeted, in the form of an implantable device.

Standard practice is to answer the above questions at a multidisciplinary surgical conference, where IEEG, brain imaging, and supportive metadata are reviewed manually. Unfortunately, the outcomes of these discussion vary from center to center, making it difficult to aggregate data across institutions for retrospective analysis and prospective clinical trials. A number of groups, including ours, are working on quantitative methods to guide therapy, but to our knowledge, none have yet performed strongly enough or been validated in prospective, large-scale trials to be used in routine clinical practice[18,19]. An important publication by Astner-Rohracher et al. recently suggested the need to compare new methods for guiding therapy against clinical predictions based upon the pre-implant data alone[20]. The group's "5-Sense Score" was able to predict seizure focality in patients prior to invasive electrode implantation with an AUC of 0.67. This modest but important result validates the need to proceed with IEEG recording in these patients, is easy to apply across centers, and provides a floor of accuracy (null model) against which additional methods should be benchmarked to demonstrate value.

In emphasizing quantitative measures, our group and others hope to give clinicians rigorous, reliable tools to aid in clinical decision making, to improve our understanding of human epileptogenic networks, and to improve standardization across centers that will facilitate large-scale, multi-center clinical trials of existing and new invasive therapies as they arise[14,18,19]. As quantitative methods are typically compared to clinical experts, we chose to compare our results to expert determinations of seizure focality, choice of therapy, and patient outcome.

Clinicians localize epileptic networks with IEEG by inducing seizures to measure their onset and spread patterns (ictal recording). Although interictal or seizure-free IEEG data are not routinely used to classify epilepsy during surgical evaluation, quantitative analysis of interictal segments has proven valuable for localization.[21–25] Additionally, exclusively using ictal data for localization in the acute inpatient setting may be misleading.[26,27] While there have been several attempts to use quantitative interictal IEEG biomarkers to predict outcome following epilepsy surgery, using these measures to look at the simpler question of whether these networks are focal or distributed remains unexplored. Addressing this question may obviate the need for ictal recording if the decision is between focal intervention or more central neuromodulation.

## Methods

In this study, we aimed to determine whether a quantitative analysis of interictal IEEG data could predict the focality of epileptic networks on IEEG better than pre-implant estimates. We hypothesized that the spatial dispersion of abnormalities, quantified through interictal IEEG using a normative atlas, would correlate with the focality of epileptic networks.[21,28] To test this hypothesis, we classified patients into either focal or non-focal groups based on clinical assessments of IEEG focality. The focal group comprised patients with a focal seizure-onset zone on IEEG, while the non-focal group included those with bilateral, multifocal, broad, or non-localized seizure onset zones. We incorporated the 5-SENSE Score, which summarizes noninvasive clinical variables to predict focality,[29] and implemented a measure quantifying the spatial density of both the implanted IEEG electrodes as well as their associated abnormalities. We then used our calculated measures to predict ultimate therapy and surgical outcome, hypothesizing that poor outcomes and palliative therapies are more common in patients with non-focal epileptic networks.

### Patient information

We analyzed 101 drug-resistant epilepsy patients from The Hospital of the University of Pennsylvania who underwent pre-surgical evaluation and IEEG. Patients were enrolled serially between 2011 and 2021 after providing written informed consent for IEEG data analysis, in line with the University of Pennsylvania's IRB-approved protocol (reference number 821778). Patient characteristics are shown in **Table 1**.

### Clinical data extraction

**Non-invasive pre-surgical evaluation phase:** We curated information from each patient's written clinical notes on: (1) scalp EEG seizure laterality and localization, (2) scalp EEG spike laterality and localization, (3) MRI lesion status, laterality, localization, and type (4) neurophysiological testing for language dysfunction lateralization, (5) semiology lateralization and localization. These non-invasive clinical variables aided in formulating hypotheses about the localization and lateralization of epileptic networks.[29] We then summarized these variables using the recently proposed 5-SENSE score[29], which combines five pre-implant clinical variables that were found to predict SEEG focality: a) extent of the lesion on magnetic resonance imaging, b) extent of the ictal discharge, c) extent of interictal epileptiform discharges (IEDs), d) strength of localizing semiology, and e) localization of neuropsychological deficit. More details on 5-SENSE score are available elsewhere[29].

**Invasive IEEG evaluation phase:** Following the previous non-invasive phase, IEEG electrodes were implanted to target the epileptic network. Patients were admitted to EMU for IEEG implantation, and seizures were induced to determine the laterality and localization of seizure focus. From discharge summaries and multidisciplinary conference notes, we determined whether the IEEG-defined seizure onset zone was classified as focal, bilateral, multifocal, broad, non-localized, or missed. Based on this data, 65 patients were focal and 36 were non-focal. The non-focal group included those with bilateral, multifocal (2 or more discrete onset zones), broad (lobar or multi-lobar), non-localized, or missed seizure onset zones.

**Post-implant therapy selection, surgery, and follow-up phases:** The choice of therapy—resection, ablation, or stimulation—was determined by weighing the risk-benefit ratio between achieving seizure freedom and potential neurological or cognitive deficits. Generally, patients found to have focal seizure onset on IEEG received surgery, and patients categorized as non-focal received neuromodulatory devices, but exceptions were seen: patients with suspicion for multifocal seizures underwent surgery if a key node was thought to be mediating, and focal patients underwent device implantation if seizure onset occurred in the eloquent cortex. Post-surgical ILAE outcomes were derived at 2 years. Table 1 summarizes the surgical outcome and therapy data.

## IEEG electrode reconstruction on MRI

We implemented iEEG-recon, a modular software tool, to reconstruct IEEG electrode contacts on pre-implant T1-weighted MRI images.[30] This tool enables automatic labeling, registration, and reconstruction of iEEG electrode coordinates on brain images. In brief, each electrode contact was semi-automatically labeled on CT images, which were then registered to pre-implant T1-weighted MRI. Brain segmentation and parcellation were conducted in each subject's native space using Freesurfer.[30] Electrodes were classified as being in grey matter, white matter, or outside the brain. We excluded all electrodes located in white matter and outside the brain. For all grey matter electrodes, we determined their corresponding regions based on the Desikan-Killiany parcellation scheme in the patient's native-space.

## IEEG data processing

We selected 20 one-minute interictal IEEG clips for each patient's stay in the Epilepsy Monitoring Unit.[31] These clips were randomly chosen, excluding three days post-implantation and two hours from any seizure annotation to minimize implant effects and peri-ictal changes during awake periods. To detect awake periods in iEEG, we applied an alpha/delta power ratio (ADR, normalized threshold = -0.40) detector, previously described for identifying the highest probability of wakefulness.[32] We obtained these clips from IEEG.org and applied bipolar referencing. The data was processed using a 3rd order Butterworth bandpass filter between 0.5 and 80Hz, a 60Hz notch filter, and down sampled to 200Hz. We calculated relative band power (IEEG spectral activity) and magnitude squared coherence (IEEG connectivity) in 5 canonical bands (delta 0.5-4Hz, theta 4-8Hz, alpha 8-12Hz, beta 12-30Hz, gamma 30-80Hz), as well as total broadband (0.5-80Hz) power and coherence, using a 2s Hamming window with a 1s overlap. These preprocessing steps are identical to those in our previous work[21], with the only exception being that we selected 20 interictal IEEG clips and performed the analysis on all 20 clips to ensure temporal robustness of our quantitative methods.

## Quantifying interictal IEEG abnormalities from normative IEEG atlas

Normative modeling is an innovative case-control approach that quantifies abnormalities in a patient's IEEG signals by determining how deviant they are from the normal range expected in controls. As IEEG is rarely implanted in patients without epilepsy, we created a normative IEEG atlas by concatenating 2304 electrodes implanted outside epileptogenic tissues, which were presumed healthy, across 144 patients. This normative IEEG atlas has been described in our previous work and is publicly available.[21]

We quantified interictal IEEG abnormalities from both IEEG spectral activity and connectivity. Briefly, for each IEEG contact's relative band power in each canonical frequency band, we calculated the Z-score from the electrodes in the same parcellated ROI in the normative IEEG atlas. Similarly, for iEEG coherence connectivity between two ROIs in each frequency band, we computed the Z-score of equivalent connections between the same ROIs in the normative iEEG atlas. Z-score represents the deviation from normal IEEG spectral activity and coherence. We repeated these steps for all 20 one-minute interictal clips, computing a distribution of Z-scores for iEEG activity and connectivity. These steps are identical to those in our previous work and are explained in detail there.[21]

## Quantifying spatial coverage of IEEG and spatial dispersion of IEEG abnormalities

The pre-implant clinical hypothesis regarding the focality of the epileptic network influences the spatial coverage of the IEEG implant. In cases where epileptic networks are likely focal, the density of IEEG may be more concentrated in areas believed to be involved in seizure generation and propagation. Conversely, if clinicians suspect a non-focal epileptic network, the IEEG coverage is likely to be broader.

To quantify this clinical intuition about the focality of the epileptic network, we computed standard distance between IEEG contacts. Standard distance (SD) represents the 3-D standard deviation of each patient's IEEG electrode coverage (Equation 1).[33] Mathematically, if an IEEG contact *i* has coordinates (x, y, z) and there are n contacts in a patient, the standard distance can be computed as follows:

$$SD = \sqrt{\frac{\sum_{i=1}^{n}(x_i - \bar{X})^2}{n} + \frac{\sum_{i=1}^{n}(y_i - \bar{Y})^2}{n} + \frac{\sum_{i=1}^{n}(z_i - \bar{Z})^2}{n}}$$

A high SD value indicates a wider coverage, while a smaller SD value represents a more compact coverage. We refer to the standard distance between IEEG contacts as the "implant distance".

To compute the spatial dispersion of IEEG abnormalities, we implemented weighted standard distance (WSD). Weighted standard distance weights the SD with the abnormality values of each iEEG contact. Mathematically, if the abnormality of an IEEG contact *i* with coordinates (x, y, z) is w, and there are n contacts in a patient, the weighed standard distance can be computed as follows:

$$WSD = \sqrt{\frac{\sum_{i=1}^{n} w_i (x_i - \bar{X}_w)^2}{\sum_{i=1}^{n} w_i} + \frac{\sum_{i=1}^{n} w_i (y_i - \bar{Y}_w)^2}{\sum_{i=1}^{n} w_i} + \frac{\sum_{i=1}^{n} w_i (z_i - \bar{Z}_w)^2}{\sum_{i=1}^{n} w_i}}$$

where $\{\bar{X}_w, \bar{Y}_w, \bar{Z}_w\}$ represent the weighted mean center:

$$\bar{X}_w = \frac{\sum_{i=1}^{n} w_i x_i}{\sum_{i=1}^{n} w_i}$$

In simpler terms, the weighted standard distance (WSD) takes into account the abnormality value at each IEEG contact and measures the spatial distribution of abnormality. A high WSD value indicates that abnormalities on IEEG are spatially widespread, while a low WSD value suggests that the abnormalities are located spatially closer together. We refer to the weighted standard distance of IEEG abnormalities as the "abnormality distance". The concepts described above are illustrated schematically in the overview presented in **Figure 1**.

## Statistics
We conducted univariate analysis to test how each metric discriminated focal from non-focal iEEG-seizure onset. In these univariate tests, we calculated effect sizes using the Cohen's d score, non-parametric effect sizes using the area under the receiver operating characteristic curve (AUC), and non-parametric two-tailed p-values from the Mann-Whitney U test. We combined these quantitative metrics using multivariate logistic regression with lasso regularization, to mitigate the impact of correlated features and reduce the number of features. We applied leave-one-out cross-validation to predict IEEG focality in unseen (test) patients. To compare different predictive models, we applied the DeLong's test. We performed all analyses with Python 3.10, and default regularization parameters for logistic regression ('l1' norm, C=1) in sklearn.

## Data Availability Statement

All IEEG data used in this study are publicly available at https://www.ieeg.org/. The normative iEEG atlas can be accessed at: https://discover.pennsieve.io/datasets/179. The iEEG-recon software is available at https://ieeg-recon.readthedocs.io. Documented codes and the 5-SENSE score for all patients included in our analysis can be found at: [GitHub link].

# Results

101 patients met the inclusion criteria. Clinicians determined that seizure generators were focal in 65 patients and non-focal in 36 patients. Among the 36 non-focal patients, 11 underwent surgery and 15 received stimulation therapy. Out of the 65 focal patients, 52 had surgery and 7 were treated with stimulation therapy. **Table 1** provides an overview of surgical outcomes and therapy information. Our primary research questions were 1) Can quantifying abnormalities in interictal IEEG differentiate between focal and non-focal epileptic networks? and 2) How well can interictal quantitative measures predict choice of therapy and patient outcome?

## Quantitative features distinguish focal and non-focal epileptic networks

We conducted univariate analyses on: a) 5-SENSE score, b) spatial coverage of IEEG (implant distance), and c) spatial dispersion of abnormalities quantified from interictal IEEG activity and connectivity features (abnormality distance). We investigated the association of these variables to determine IEEG focality. As depicted in **Figure 2**, these variables significantly distinguish patients with focal epileptic networks from those with non-focal networks. Among these variables, as shown in Figure 2c-d, the 5-SENSE score had the highest effect size ($p = 0.004$, $d = 0.72$, AUC = 0.68), followed by the abnormality distance in gamma band power ($p = 0.004$, $d = 0.69$, AUC = 0.68).

Although these variables could independently distinguish between focal and non-focal patients, they exhibited a weak correlation to the 5-SENSE, as shown in Figure 2a-b. In particular, the correlation between the 5-SENSE score and the implant distance was $r = -0.22$ ($p = 0.024$), implying that patients with a higher pre-implant probability of focality had a less spatially-dispersed implant. The correlation between the 5-SENSE score and the abnormality distance in the gamma band was $r = -0.13$ ($p = 0.19$), implying a non-significant trend toward lower gamma band abnormality dispersion in patients with a higher pre-implant probability of focality. The combination of low correlation among variables and their individual discriminatory ability in determining focal and non-focal epileptic network suggests that these variables may have complementary information when combined.

## Predicting focality by combining preimplant and interictal IEEG data

To predict IEEG focality, we conducted cross-validated logistic regression classification using three distinct models: a) model incorporating only the 5-SENSE scores, b) model combining features from 5-SENSE scores and implant distance, and c) model combining features from 5-SENSE scores and IEEG abnormality distances.

Figure 3 shows the comparison of these three models, along with the prediction performance for each. We found that the combination of 5-SENSE scores and interictal IEEG abnormality distances predicts IEEG focality most accurately, with a cross-validated AUC of 0.79 (Fig. 3b-c). In contrast, the 5-SENSE score and its combination with implant distances achieved cross-validated AUCs of 0.67 and 0.68, respectively. IEEG abnormality distances were amongst the most important features (Fig. 3a). The combined model predicted focality with specificity of 86% at the optimal operating point of the ROC curve (Fig. 3d).

**Focality prediction relates to surgical outcomes and implantation types**

We next examined whether the output of our model, trained to detect focality from interictal iEEG data, would also vary by surgical outcome, implantation type, or subtypes of non-focal epilepsy.

Figure 4a shows that the model combining 5-SENSE score and IEEG predicted a higher probability of a focal epileptic network in patients who achieved seizure freedom post-surgery, versus those who continued to experience seizures after surgery and those who underwent neurostimulation (AUC = 0.81, $p < 0.05$). This suggests that more focal networks are associated with better surgical outcomes. In terms of implant type, Figure 4b shows that the effect size between focal and non-focal epileptic networks was greater in patients with ECOG implants (AUC = 0.93) compared to those with SEEG implants (AUC = 0.73). Amongst subgroups of 'non-focal' patients, Figure 4c shows that quantitative prediction of focality significantly differentiate both patients with bifocal seizure onsets from those with unifocal seizure onset (AUC = 0.85) and patients with multifocal onsets from those with unifocal onset (AUC = 0.74). There was no significant difference in the quantitative prediction of focality between bifocal and multifocal patients.

Figure S1 shows consistent equivalent results as in Figure 4a for seizure outcomes measured on Engel scale and for different predictive models. Figure S2 shows consistent equivalent results as in Figure 4b-c for different choices of predictive models. Collectively, these findings provide additional validation of our approach, indicating that the predicted probability of a focal epileptic network tends to be lower in cases where therapies are palliative, or when the IEEG has a broader spatial coverage.

## Discussion

In this study we present a quantitative method to categorize the focality of epileptic networks for individual patients based upon their IEEG, by comparing their data to an atlas of prior patients. The method combines a) pre-implant clinical variables, b) spatial coverage of IEEG implant, and c) spatial dispersion of abnormalities quantified on interictal IEEG segments. Our findings suggest that we can combine these variables to predict whether a patient has a focal or a distributed epileptic network. These findings have significant clinical implications. Patients with a high probability of having a focal epileptic network might be better candidates for focal surgical intervention, such as laser ablation or surgical resection, and require more complex analyses, such as from ictal recordings. Conversely, patients with distributed or multifocal epileptic networks, and currently a lower likelihood of seizure-free outcome, might benefit more from palliative neurostimulation therapies that reduce seizure frequency and/or severity, but with a much lower chance of rendering them seizure-free. In these cases, our method might suggest that clinicians consider regional or central neurostimulation therapy earlier, rather than "chasing" poorly localized ictal onset patterns. Since our method does not rely on ictal recording, it has the potential to reduce length of stay during implantation, reduce morbidity from seizures associated with medication withdrawal, and reduce cost in patients found unlikely to have a clear focal onset.

Of interest, our computation of the 5-Sense Score replicated the findings of the original study. This reinforces that study's generalizability to other tertiary care centers. In an interesting confirmatory result, we found that a quantitative measure of the spatial dispersion of implanted IEEG electrodes predicted network focality with similar precision as the 5SS, indicating that in our center the implant strategy accurately translated the preimplant data into the next phase of presurgical evaluation. We found that a model incorporating preimplant data with the spatial dispersion of IEEG interictal abnormality improved significantly upon estimates of whether a given patient's network demonstrates a focal or non-focal seizure onset pattern.

Perhaps unsurprisingly, given the relationship between the clinical definition of focality and the decision to recommend focal surgery vs. neuromodulation via an implantable device (Table 1), we found that models of focality were effective in predicting whether patients were referred to resective surgery or ablation, or for device therapy (Figure 4). Generally, 'focal' patients received surgery and 'nonfocal' patients received neuromodulatory devices in our cohort though exceptions exist clinically; for example patients with focal seizure onset in eloquent cortex or in regions anatomically excluded from resection (e.g. because of vascularity or access issues) may receive neuromodulation.[34] In addition, patients with less localized seizures may have received surgery if key nodes were thought to be most implicated, as inferred from seizure spread or for other reasons.[12,13] While our model does not include these more advanced considerations, it does provide a framework for including other quantitative measures that could, and it also provides one benchmark for assessing whether these more subjective decisions improved outcome more than quantitative measures of focality alone.

In addition, we observed that quantitative models of focality could differentiate between favorable and unfavorable surgical outcomes (Figure 4). All models (5-SENSE, 5-SENSE + Implant distance, combined 5-SENSE + EEG) discriminated ILAE 1 or 2 from ILAE 3+. However, only the IEEG model discriminated Engel 1 from Engel 2+ outcomes. Presumably, broader epileptogenic networks may be less technically amenable to ablation or resection therapy, but an exact mechanism by which the spatial dispersion of abnormality relates to outcomes in surgical patients also remains unclear, and it is likely a function of the specific physiology of each patient.[10,12,13,35,36] Interestingly, abnormality distances both in relative power/coherence bands and in absolute broadband power were found to be significant predictors (Supplemental Tables 1,2,3) in a logistic regression model. Nonlinear interactions across frequency bands may play a role in the focality of seizure onset zones,[37–39] which would likely require larger datasets to disentangle with more complex models.

The practice of our center over the time course of this study reflects wider trends in North America toward more SEEG implantation and less ECoG.[40] We observed that all models performed better in discriminating focal from nonfocal networks in patients implanted with subdural ECoG electrodes compared to patients implanted with SEEG, including the 5-SENSE Score, which was developed on an entirely SEEG population. This phenomenon may reflect selection bias surrounding which patients were chosen for subdural ECoG implants vs SEEG implants: Subdural ECoG implants are often used to delineate seizure onset regions from eloquent cortex, often around focal lesions, whereas patients undergoing SEEG often receive broad cortical and subcortical sampling, potentially reflecting a more diffuse pre-implant hypothesis.

The need to localize seizure onset and spread, beyond the baseline clinical hypothesis or our analysis of background interictal abnormality, warrants further study and rigorous quantification to incorporate these measures other biomarkers of epileptogenicity with our interictal and implant spatial measures. The literature here is more limited, relating onset region and time of spread to other regions, and its impact on eventual outcome are best documented for medial temporal lobe epilepsy. For example, a study by Andrew et al. demonstrates that seizures that spread rapidly outside the resection margin associated with a poorer outcome after temporal lobectomy[41]. The literature is less clear for seizures that spread around focal lesions and to other structures with strong connectivity, for example seizures that spread rapidly from mesial temporal lobe to insula. The additional degrees of freedom in these types of analyses will require a much larger number of patients in an atlas of ictal recordings, interventions, and known outcomes in order to quantify and understand this process.

An advantage of the interictal method presented in the current study is its much more limited application and use. The method requires only 20 minutes of interictal IEEG data to map focality. Should a non-focal network be predicted, this could obviate the need for prolonged ictal recording and facilitate more expeditious neuromodulation treatment. Increasing literature measuring interictal brain abnormalities[21,22,42,43] and perturbing networks with stimulation[44–46] may lead intracranial studies to rely less on eliciting spontaneous seizures in the future to identify the major areas of dysfunction in epileptic networks amenable to treatment. It is likely that some combination of these approaches, with quantitative metrics of seizure dynamics, incorporating data on brain regions of interest and spatial distance, will likely be needed to improve on the present model. Central to this improvement is the need to aggregate and collaboratively analyze data from many patients in many centers using standardized protocols. We hope that using large numbers of patients can compensate for sparse electrode sampling in each patient.

Ultimately, predictive computational models of epileptogenicity may help guide surgical decision-making. In this study, we related the IEEG activity and connectivity of each patient to a corpus of prior patients. We show spatial coverage of electrodes, their associated IEEG abnormalities, and the data used in determining that coverage relates to seizure onset extent and surgical outcomes. By this framework, future studies may elucidate the cost-benefit decision analysis of implanting each additional electrode to identify seizure generators, or perhaps obviate the need to move forward with ictal recording in cases where more diffuse network topology is likely.

This study has important limitations. As a single-center, retrospective cohort study at a tertiary care institution, generalizability to other centers has not yet been determined. We used cross-validated lasso logistic regression to assess out-of-sample performance. There is considerable heterogeneity across epilepsy patients, and while a cohort of 101 patients is relatively large for such studies of IEEG, studies across multiple centers with greater numbers of patients are likely necessary to capture the true variability in 'non-focal' onset. The use of normative IEEG by identifying uninvolved epileptic electrodes is a nascent field. The openly-available atlas we employed likely requires substantially more data to improve the accuracy and generalizability of the models. We do make assumptions that unaffected electrodes in patients with epilepsy reflect true 'normal' activity and connectivity. This assumption is likely incorrect, as they are derived from epileptic brains, which likely skews the definition of 'normal.' However, the results of this study and others in normative IEEG modeling show promise in providing measures of epileptogenicity and support the continued use of this approach, for now, and contribution to a central collaborative corpus of IEEG data. Finally, it will be important for future studies of interictal IEEG to incorporate other biomarkers of epileptogenicity besides band power and connectivity, such as interictal spikes and high frequency oscillations.[47–50]

In conclusion, in this study we propose rigorous, data-driven quantitative measures to assess the spatial extent of epileptic networks in the brain. The quantitative methods we tested can complement presurgical evaluations to decrease lengthy hospital stays and optimize patient and therapy selection in drug-resistant epilepsy. Ultimately, these tools aim to standardize clinical decisions in ways that can be deployed in multi-center studies to advance and optimize care.

**Funding:**

RSG was supported by the National Institute of Neurological Disorders and Stroke of the National Institutes of Health under Award Number T32 NS-091006. NS received support from American Epilepsy Society (953257), NINDS (R01-NS-116504, R01-NS-125137). Erin Conrad received support from the National Institute of Neurological Disorders and Stroke (NINDS; R25 NS-065745, K23 NS121401-01A1) and the Burroughs Wellcome Fund. Dr. Litt is supported in this research by DP1-NS-122038, R01-NS-125137, The Mirowski Family Research Fund, Neil and Barbara Smit and Jonathan Rothberg. Disclosures/Competing Interests: Erin Conrad consults for Epiminder, an EEG device company.

**Acknowledgements:**

We thank Carolyn Wilkinson, Jacqueline Boccanfuso, Magda Wernovsky and all other members and staff of the Center for Neuroengineering and Therapeutics for assistance with data organization.

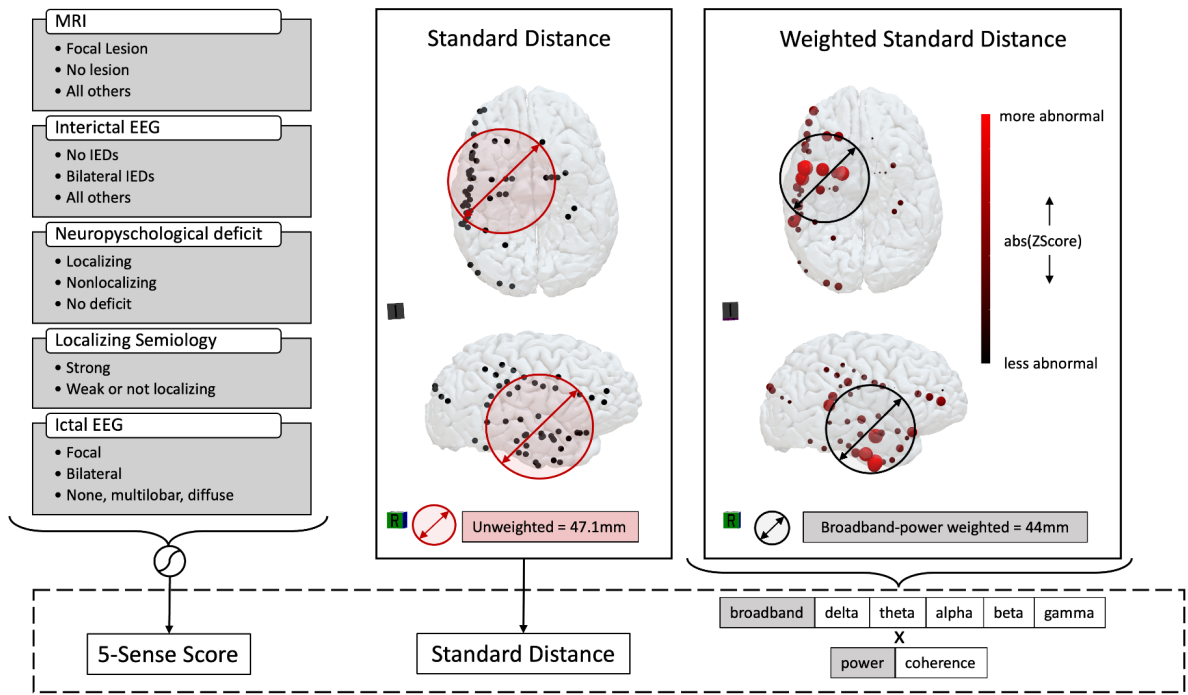

**Figure 1.** Features calculated to discriminate patients found to have focal from non-focal seizure onset with IEEG. Left: 5-SENSE score using preimplantation data previously demonstrated to have predictive value (IED = Interictal epileptiform discharge). Middle: Standard Distance (*implant distance*), calculated as the standard deviation of distance from each electrode to the mean center of the electrodes in space. Right: Weighted Standard Distance (*abnormality distance*), calculated as the standard deviation of the dispersion of iEEG abnormality for each of 6 bands in both band-power and coherence.

**Table 1.** Patient Characteristics

|  | **Non-focal (n=36)** | **Focal (n=65)** | **p*** |
|---|---|---|---|
| Female, n | 22 | 30 | 0.17 |
| Male, n | 13 | 35 |  |
| Age of onset, mean (std) | 17.6 (12.5) | 16 (12.4) | 0.53 |
| Age at implant, mean (std) | 36.6 (10.6) | 35.5 (12.2) | 0.50 |
| MRI lesional, n | 17 | 42 | 0.13 |
| Focal lesion | 5 | 28 | 0.02 |
| Nonfocal lesion+ | 12 | 14 |  |
| Gray matter channels, mean (std) | 45.3 (14.8) | 37.82 (17) | 0.06 |
| Grids/strips/depths, n | 8 | 26 | 0.13 |
| SEEG | 27 | 39 |  |
| Surgery, n | 11 | 52 | <0.001** |
| Ablation | 6 | 24 | 0.82 |
| Resection | 5 | 29 |  |
| Surgical outcomes (evaluated at 2 years) |  |  |  |
| ILAE 1-2 | 5 | 28 | 1.0 |
| ILAE 3-6 | 3 | 16 |  |
| Neurostimulation device, n | 15 | 7 |  |

*chi-square for categorical data, U test for continuous; **comparison of surgery vs. device; +non-focal MRI lesion includes multifocal, broad, and bilateral lesions[29]

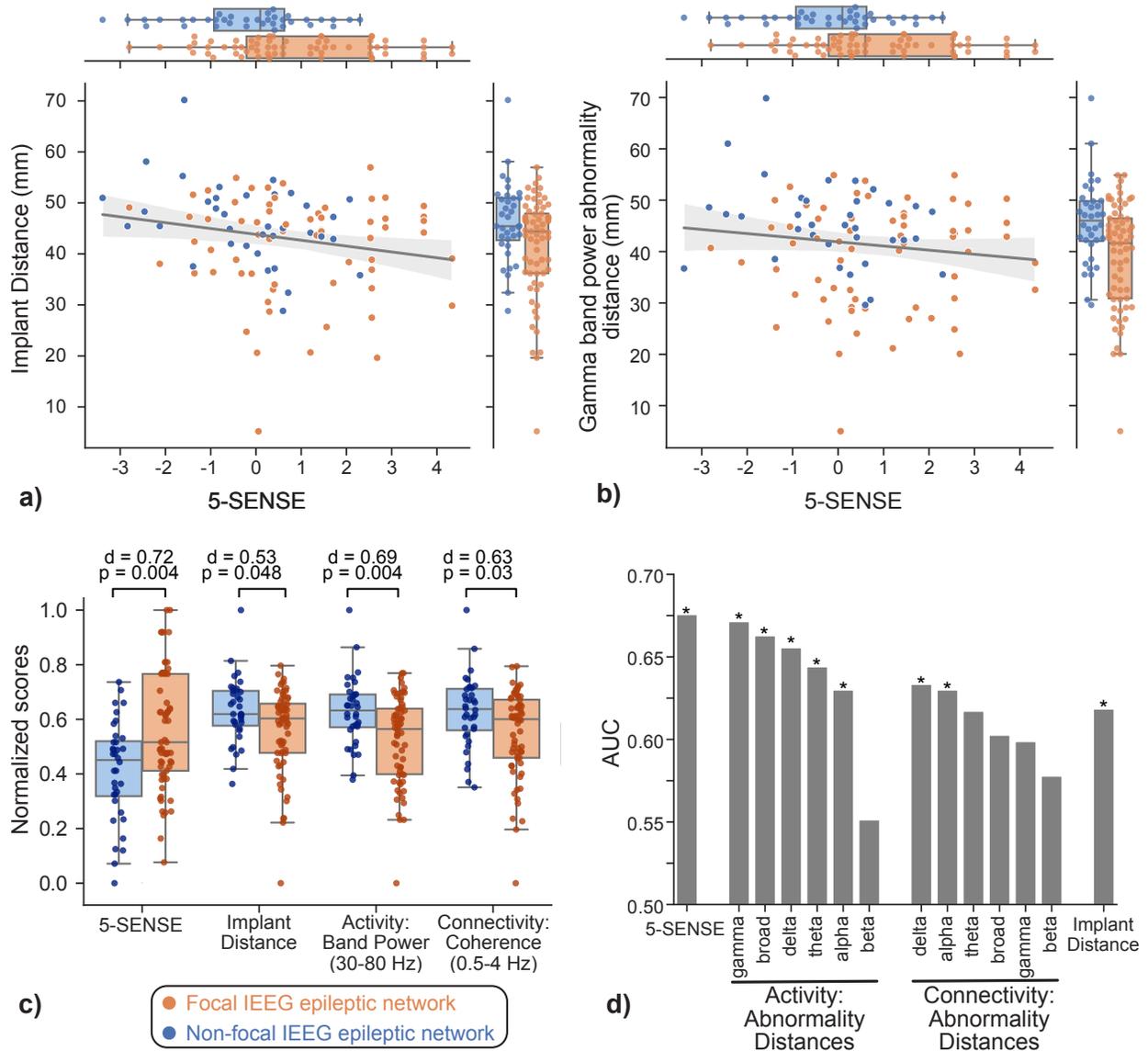

**Figure 2. Determining focality of epileptic networks from quantitative preimplant and IEEG features.** Each dot in the scatter plots represents a patient. a) Implant distance and 5-SENSE score discriminate patients with focal and non-focal epileptic networks, but they have a weak negative correlation (r = -0.22, p = 0.024) with each other. b) Spatial dispersion of IEEG spectral abnormalities in the gamma band can differentiate focal and non-focal epileptic networks; these IEEG abnormality features are not correlated with 5-SENSE score (r=-0.16, p=0.11). c) Individually, the 5-SENSE score, implant distance, and spatial dispersion of IEEG abnormalities distinguish focal and non-focal epileptic networks. d) Bar plot showing the discrimination power of each quantitative feature in differentiating focal and non-focal epileptic networks.

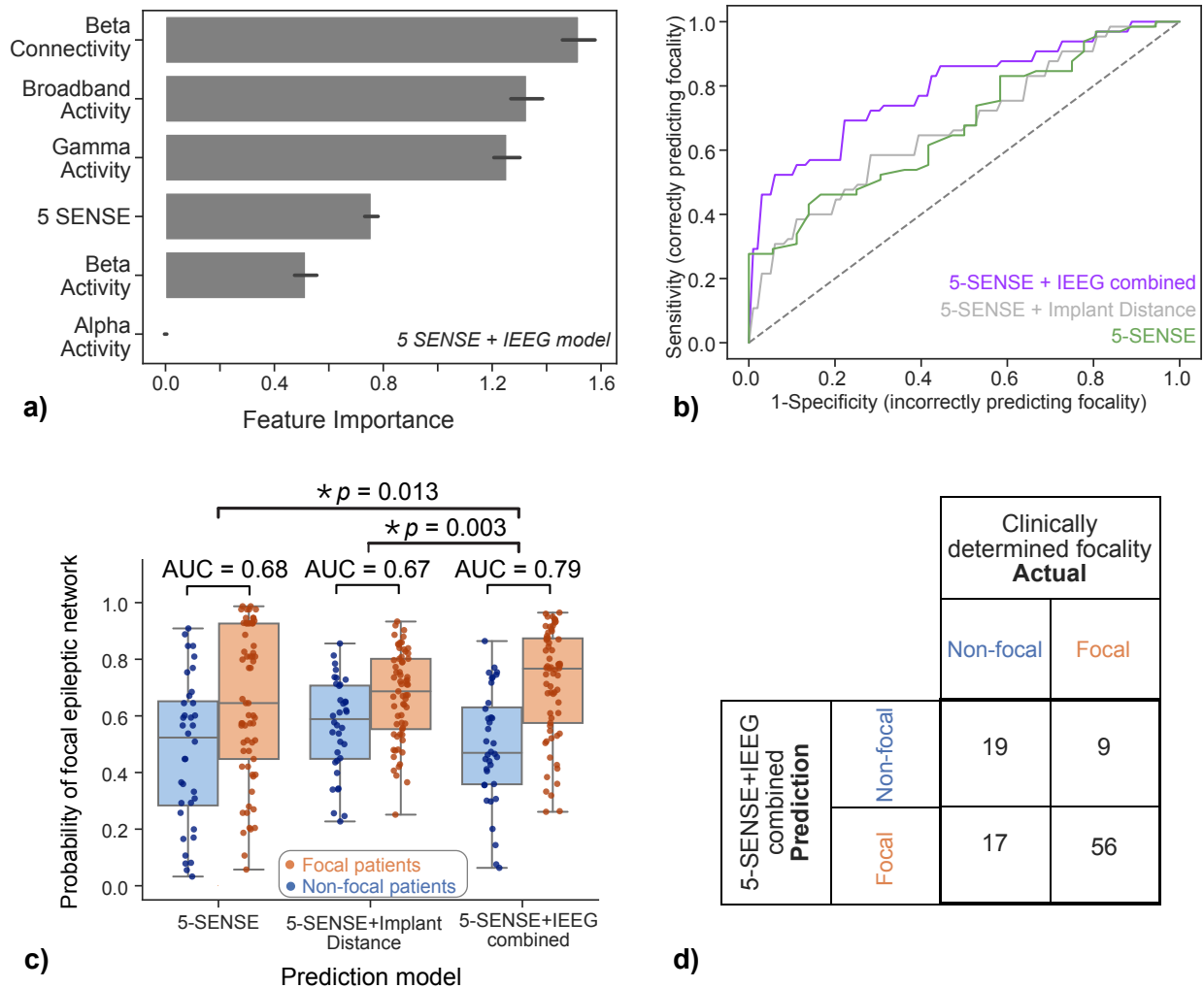

**Figure 3. Multivariable model combining IEEG features with the 5-SENSE score to predict IEEG focality.** a) Features combine in a linear model resulting in AUC=0.79 to predict focality. Mean ± SD of feature coefficients across cross-validation folds with ridge regularization b) Comparison of ROC curves of the predictions of the 5SENSE+IEEG model (blue) to a preimplant baseline model incorporating only 5-SENSE+implant distance (grey) and the 5-SENSE score (green). c) 5-SENSE scores combined with interictal IEEG abnormality distances predict IEEG focality most accurately (cross-validated AUC 0.79), outperforming 5-SENSE score alone and its combination with implant distances (cross-validated AUCs 0.67 and 0.68, respectively), as demonstrated by DeLong's test p-values on the box plots. D) Confusion matrix of the model incorporating IEEG features and the 5-SENSE score.

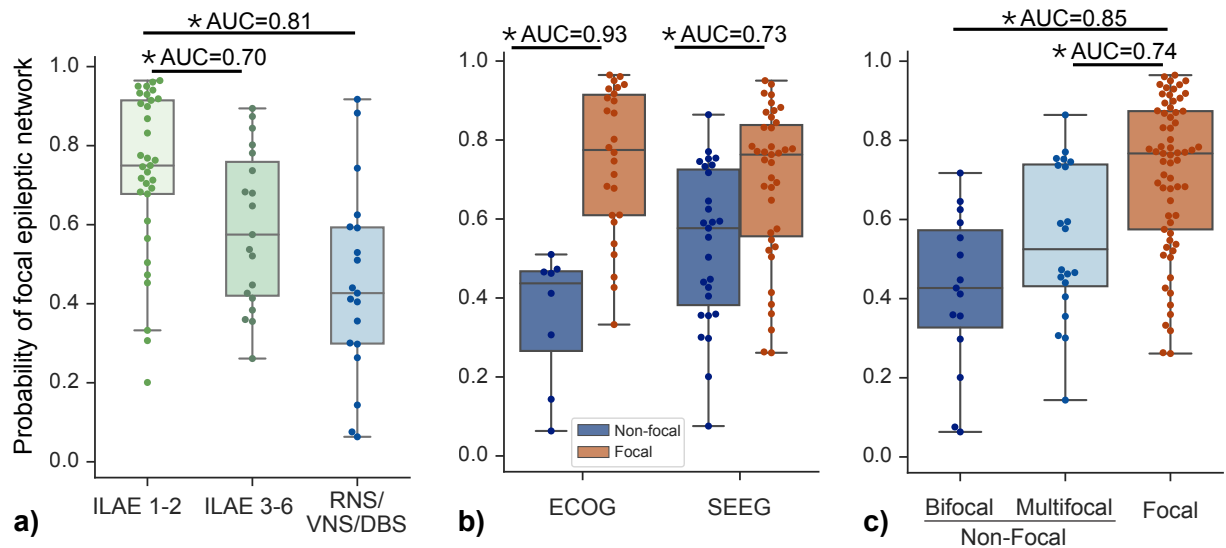

**Figure 4: Predictive value of the 5-SENSE model combined with IEEG for focality and its implications for surgical outcomes and implantation types. (a)** Model indicates a higher likelihood of focal epileptic networks in patients achieving post-surgery seizure freedom (AUC=0.81, p<0.05). **(b)** Greater effect size in focal vs. non-focal networks for patients with ECOG implants (AUC=0.93) than SEEG implants (AUC=0.73). **(c)** Among 'non-focal' patients, focality prediction significantly differentiates bifocal from unifocal (AUC = 0.85) and multifocal from unifocal epileptic networks (AUC = 0.74), with no significant difference between bifocal and multifocal patients.

Reference:

1. Kwan P, Arzimanoglou A, Berg AT, et al. Definition of drug resistant epilepsy: Consensus proposal by the ad hoc Task Force of the ILAE Commission on Therapeutic Strategies. *Epilepsia*. 2010;51(6):1069-1077. doi:10.1111/j.1528-1167.2009.02397.x

2. Kwan P, Schachter SC, Brodie MJ. Drug-Resistant Epilepsy. *New Engl J Medicine*. 2011;365(10):919-926. doi:10.1056/nejmra1004418

3. Wiebe S, Blume WT, Girvin JP, Eliasziw M, Group E and E of S for TLES. A Randomized, Controlled Trial of Surgery for Temporal-Lobe Epilepsy. *New Engl J Medicine*. 2001;345(5):311-318. doi:10.1056/nejm200108023450501

4. Engel J, Wiebe S, French J, et al. Practice parameter: Temporal lobe and localized neocortical resections for epilepsy Report of the Quality Standards Subcommittee of the American Academy of Neurology, in Association with the American Epilepsy Society and the American Association of Neurological Surgeons. *Neurology*. 2003;60(4):538-547. doi:10.1212/01.wnl.0000055086.35806.2d

5. Jobst BC, Cascino GD. Resective Epilepsy Surgery for Drug-Resistant Focal Epilepsy: A Review. *Jama*. 2015;313(3):285-293. doi:10.1001/jama.2014.17426

6. Vakharia VN, Duncan JS, Witt J, Elger CE, Staba R, Engel J. Getting the best outcomes from epilepsy surgery. *Ann Neurol*. 2018;83(4):676-690. doi:10.1002/ana.25205

7. Sinha N, Dauwels J, Kaiser M, et al. Predicting neurosurgical outcomes in focal epilepsy patients using computational modelling. *Brain*. 2017;140(2):319-332. doi:10.1093/brain/aww299

8. Sinha N. Localizing epileptogenic tissues in epilepsy: are we losing (the) focus? *Brain*. 2022;145(11):3735-3737. doi:10.1093/brain/awac373

9. Bernabei JM, Li A, Revell AY, et al. Quantitative approaches to guide epilepsy surgery from intracranial EEG. *Brain*. Published online 2023. doi:10.1093/brain/awad007

10. Bartolomei F, Lagarde S, Wendling F, et al. Defining epileptogenic networks: Contribution of SEEG and signal analysis. *Epilepsia*. 2017;58(7):1131-1147. doi:10.1111/epi.13791

11. Andrews JP, Gummadavelli A, Farooque P, et al. Association of Seizure Spread With Surgical Failure in Epilepsy. *Jama Neurol*. 2019;76(4):462-469. doi:10.1001/jamaneurol.2018.4316

12. Spencer DD, Gerrard JL, Zaveri HP. The roles of surgery and technology in understanding focal epilepsy and its comorbidities. *Lancet Neurology*. 2018;17(4):373-382. doi:10.1016/s1474-4422(18)30031-0

13. Barba C, Rheims S, Minotti L, et al. Temporal plus epilepsy is a major determinant of temporal lobe surgery failures. *Brain*. 2016;139(2):444-451. doi:10.1093/brain/awv372

14. Sinha N. Localizing epileptogenic tissues in epilepsy: are we losing (the) focus? *Brain J Neurology*. 2022;145(11):3735-3737. doi:10.1093/brain/awac373

15. Sinha N, Dauwels J, Kaiser M, et al. Predicting neurosurgical outcomes in focal epilepsy patients using computational modelling. *Brain*. 2017;140(2):319-332. doi:10.1093/brain/aww299

16. Sinha N, Wang Y, Silva NM da, et al. Structural Brain Network Abnormalities and the Probability of Seizure Recurrence After Epilepsy Surgery. *Neurology*. 2021;96(5):e758-e771. doi:10.1212/wnl.0000000000011315

17. Sinha N, Peternell N, Schroeder GM, et al. Focal to bilateral tonic–clonic seizures are associated with widespread network abnormality in temporal lobe epilepsy. *Epilepsia*. 2021;62(3):729-741. doi:10.1111/epi.16819

18. Sinha N, Johnson GW, Davis KA, Englot DJ. Integrating Network Neuroscience Into Epilepsy Care: Progress, Barriers, and Next Steps. *Epilepsy Curr*. Published online 2022:153575972211012. doi:10.1177/15357597221101271

19. Bernabei JM, Li A, Revell AY, et al. Quantitative approaches to guide epilepsy surgery from intracranial EEG. *Brain*. Published online 2023. doi:10.1093/brain/awad007

20. Astner-Rohracher A, Zimmermann G, Avigdor T, et al. Development and Validation of the 5-SENSE Score to Predict Focality of the Seizure-Onset Zone as Assessed by Stereoelectroencephalography. *Jama Neurol*. 2022;79(1):70-79. doi:10.1001/jamaneurol.2021.4405

21. Bernabei JM, Sinha N, Arnold TC, et al. Normative intracranial EEG maps epileptogenic tissues in focal epilepsy. *Brain*. 2022;145(6):1949-1961. doi:10.1093/brain/awab480

22. Taylor PN, Papasavvas CA, Owen TW, et al. Normative brain mapping of interictal intracranial EEG to localize epileptogenic tissue. *Brain*. 2022;145(3):939-949. doi:10.1093/brain/awab380

23. Wang Y, Sinha N, Schroeder GM, et al. Interictal intracranial electroencephalography for predicting surgical success: The importance of space and time. *Epilepsia*. 2020;61(7):1417-1426. doi:10.1111/epi.16580

24. Tomlinson SB, Porter BE, Marsh ED. Interictal network synchrony and local heterogeneity predict epilepsy surgery outcome among pediatric patients. *Epilepsia*. 2017;58(3):402-411. doi:10.1111/epi.13657

25. Lagarde S, Roehri N, Lambert I, et al. Interictal stereotactic-EEG functional connectivity in refractory focal epilepsies. *Brain*. 2018;141(10):2966-2980. doi:10.1093/brain/awy214

26. King-Stephens D, Mirro E, Weber PB, et al. Lateralization of mesial temporal lobe epilepsy with chronic ambulatory electrocorticography. *Epilepsia*. 2015;56(6):959-967. doi:10.1111/epi.13010

27. Chiang S, Fan JM, Rao VR. Bilateral temporal lobe epilepsy: How many seizures are required in chronic ambulatory electrocorticography to estimate the laterality ratio? *Epilepsia*. 2022;63(1):199-208. doi:10.1111/epi.17113

28. Frauscher B, Ellenrieder N von, Zelmann R, et al. Atlas of the normal intracranial electroencephalogram: neurophysiological awake activity in different cortical areas. *Brain*. 2018;141(4):1130-1144. doi:10.1093/brain/awy035

29. Astner-Rohracher A, Zimmermann G, Avigdor T, et al. Development and Validation of the 5-SENSE Score to Predict Focality of the Seizure-Onset Zone as Assessed by Stereoelectroencephalography. *JAMA Neurol*. 2022;79(1):70. doi:10.1001/jamaneurol.2021.4405

30. Lucas A, Scheid BH, Pattnaik AR, et al. iEEG-recon: A Fast and Scalable Pipeline for Accurate Reconstruction of Intracranial Electrodes and Implantable Devices. Published online 2023. doi:10.1101/2023.06.12.23291286

31. [2302.05734] Temporal stability of intracranial EEG abnormality maps for localising epileptogenic tissue. Accessed March 28, 2023. https://arxiv.org/abs/2302.05734

32. Conrad EC, Revell AY, Greenblatt AS, et al. Spike patterns surrounding sleep and seizures localize the seizure-onset zone in focal epilepsy. *Epilepsia*. 2023;64(3):754-768. doi:10.1111/epi.17482

33. Mitchell A, Griffin LS. *The Esri Guide to GIS Analysis, Volume 2: Spatial Measurements and Statistics, Second Edition by Andy Mitchell | Esri Press*. Vol 2. Accessed March 22, 2023. https://www.esri.com/en-us/esri-press/browse/the-esri-guide-to-gis-analysis-volume-2-spatial-measurements-and-statistics-second-edition

34. Ryvlin P, Rheims S, Hirsch LJ, Sokolov A, Jehi L. Neuromodulation in epilepsy: state-of-the-art approved therapies. *Lancet Neurology*. 2021;20(12):1038-1047. doi:10.1016/s1474-4422(21)00300-8

35. Proix T, Jirsa VK, Bartolomei F, Guye M, Truccolo W. Predicting the spatiotemporal diversity of seizure propagation and termination in human focal epilepsy. *Nat Commun*. 2018;9(1):1088. doi:10.1038/s41467-018-02973-y

36. Conrad EC, Tomlinson SB, Wong JN, et al. Spatial distribution of interictal spikes fluctuates over time and localizes seizure onset. *Brain*. 2020;143(2):554–569.

37. Ma H, Wang Z, Li C, Chen J, Wang Y. Phase–Amplitude Coupling and Epileptogenic Zone Localization of Frontal Epilepsy Based on Intracranial EEG. *Front Neurol*. 2021;12:718683. doi:10.3389/fneur.2021.718683

38. Donoghue T, Haller M, Peterson EJ, et al. Parameterizing neural power spectra into periodic and aperiodic components. *Nat Neurosci*. 2020;23(12):1655-1665. doi:10.1038/s41593-020-00744-x

39. Eissa TL, Dijkstra K, Brune C, et al. Cross-scale effects of neural interactions during human neocortical seizure activity. *Proc National Acad Sci*. 2017;114(40):10761-10766. doi:10.1073/pnas.1702490114

40. Abou-Al-Shaar H, Brock AA, Kundu B, Englot DJ, Rolston JD. Increased nationwide use of stereoencephalography for intracranial epilepsy electroencephalography recordings. *J Clin Neurosci*. 2018;53:132-134. doi:10.1016/j.jocn.2018.04.064

41. Andrews JP, Gummadavelli A, Farooque P, et al. Association of Seizure Spread With Surgical Failure in Epilepsy. *Jama Neurol*. 2019;76(4):462-469. doi:10.1001/jamaneurol.2018.4316

42. Conrad EC, Tomlinson SB, Wong JN, et al. Spatial distribution of interictal spikes fluctuates over time and localizes seizure onset. *Brain*. 2019;143(2):554-569. doi:10.1093/brain/awz386

43. Weiss SA, Staba RJ, Sharan A, et al. Accuracy of high-frequency oscillations recorded intraoperatively for classification of epileptogenic regions. *Sci Rep*. 2021;11(1):21388. doi:10.1038/s41598-021-00894-3

44. Kamali G, Smith RJ, Hays M, et al. Transfer Function Models for the Localization of Seizure Onset Zone From Cortico-Cortical Evoked Potentials. *Front Neurol*. 2020;11:579961. doi:10.3389/fneur.2020.579961

45. Hays MA, Smith RJ, Wang Y, et al. Cortico-cortical evoked potentials in response to varying stimulation intensity improves seizure localization. *Clin Neurophysiol*. 2023;145:119-128. doi:10.1016/j.clinph.2022.08.024

46. Oderiz CC, Ellenrieder N von, Dubeau F, et al. Association of Cortical Stimulation–Induced Seizure With Surgical Outcome in Patients With Focal Drug-Resistant Epilepsy. *Jama Neurol*. 2019;76(9):1070-1078. doi:10.1001/jamaneurol.2019.1464

47. Thomas J, Kahane P, Abdallah C, et al. A Subpopulation of Spikes Predicts Successful Epilepsy Surgery Outcome. *Ann Neurol*. Published online 2022. doi:10.1002/ana.26548

48. Gliske SV, Irwin ZT, Chestek C, et al. Variability in the location of high frequency oscillations during prolonged intracranial EEG recordings. *Nature Communications*. 2018;9(1):314-314. ["http://www.nature.com/articles/s41467-018-04549-2", "papers3://publication/doi/10.1038/s41467-018-04549-2"]

49. Conrad EC, Revell AY, Greenblatt AS, et al. Spike patterns surrounding sleep and seizures localize the seizure-onset zone in focal epilepsy. *Epilepsia*. 2023;64(3):754-768. doi:10.1111/epi.17482

50. Conrad EC, Tomlinson SB, Wong JN, et al. Spatial distribution of interictal spikes fluctuates over time and localizes seizure onset. *Brain*. 2019;143(2):554-569. doi:10.1093/brain/awz386